\documentclass{aip-cp}

\usepackage[numbers]{natbib}
\usepackage{rotating}
\usepackage{graphicx}
\usepackage{bm}
\usepackage{pstricks}

\newcommand{\eqb}{\begin{eqnarray}}
\newcommand{\eqe}{\end{eqnarray}}
\newcommand{\eqbn}{\begin{eqnarray*}}
\newcommand{\eqen}{\end{eqnarray*}}
\newcommand{\pdiff}[2]{\frac{\partial #1}{\partial #2}}
\newcommand{\dif}[2]{\frac{{\rm d} #1}{{\rm d} #2}}

\newcommand{\vel}{{\rm v}}

\newcommand{\mnras}{MNRAS}
\newcommand{\apj}{ApJ}

\newcommand{\aap}{A\&A}
\newcommand{\grl}{GRL}

\begin{document}


\title{On the Cosmic Ray Driven Firehose Instability}

\author[aff1]{Robyn Scott}
\author[aff1]{Brian Reville\corref{cor1}}
\author[aff2]{Anatoly Spitkovsky}

\affil[aff1]{Centre for Plasma Physics, Queen's University Belfast, Belfast BT7 1NN, Northern Ireland }
\affil[aff2]{Department of Astrophysical Sciences, Princeton University, Peyton Hall, Princeton, NJ 08544, USA}
\corresp[cor1]{Corresponding author: b.reville@qub.ac.uk}

\maketitle

\begin{abstract}
The role of the non-resonant firehose instability in conditions relevant to the precursors of supernova remnant shocks is considered.
Using a second order tensor expansion of the Vlasov-Fokker-Planck equation we illustrate the necessary conditions for the firehose to operate.
It is found that for very fast shocks, the diffusion approximation predicts that the linear firehose growth rate is marginally faster than it's resonant counterpart.
Preliminary hybrid MHD-Vlasov-Fokker-Planck simulation results using young supernova relevant parameters are presented.

\end{abstract}

\section{INTRODUCTION}

Supernova remnants offer the most likely candidate for production of the majority of Galactic cosmic-rays, with the diffusive acceleration of particles at the fast outer shocks being the 
most promising accelerating mechanism \cite{Drury}. Despite considerable advances in recent years, regarding direct observations that imply significant magnetic field amplification 
in the vicinity of the outer shocks of several nearby supernova remnants, the interplay between accelerated particles and these amplified fields remains an area of ongoing investigation.
The key issue that remains \cite{LagageCesarsky}, based on our current understanding of magnetic field amplification, is that the maximum attainable cosmic-ray energy falls short of the knee feature on the
cosmic ray spectrum at a few PeV assuming conditions relevant to the known young Galactic SNRs \cite{Bell13}. It has thus been suggested that younger, faster shocks may be the primary source
of Galactic cosmic rays at and above the knee.

Motivated by this, we re-examine the growth of linear fluctuations in the precursor of a fast parallel shock which is efficiently accelerating cosmic rays.  We focus, in these proceedings 
on the non-resonant firehose instability \cite{BlandfordFunk, Shapiro}, driven by cosmic-rays in the extended precursor of an efficiently accelerating shock.

\section{VLASOV-FOKKER-PLANCK EQUATION}

We assume that the background plasma satisfies the ideal MHD Ohm's law $\bm{E}=-(1/c)\bm{u}\times \bm{B}$, such that the electric field vanishes in the local frame.
It is thus convenient to work in a mixed coordinate frame in which particle momentum is measured in the local fluid frame, while all other quantities are measured in a fixed inertial frame.
To order $u/c$, the VFP equation thus reads
\eqb
\pdiff{f}{t}
+(\bm{u}+\bm{\vel})\cdot \bm{\nabla}f
-\left[(\bm{p}\cdot\bm{\nabla})\bm{u}\right]\cdot\pdiff{f}{\bm{p}}
-e\bm{B}\cdot\left(\bm{\vel}\times\pdiff{f}{\bm{p}}\right)=
\left(\frac{\delta f}{\delta t}\right)_{\rm c}\, ,
\eqe
where, for small angle scatterings, we take the following form for the collision operator:
\eqb
\left(\frac{\delta f}{\delta t}\right)_{\rm c}=\frac{\nu}{2}\left\lbrace\pdiff{\,}{\mu}\left[
(1-\mu^2)\pdiff{f}{\mu}\right] + \frac{1}{1-\mu^2}\frac{\partial^2 f}{\partial \phi^2}\right\rbrace\, ,
\eqe
with $\nu(p, B)$ the collision rate.

Since we seek to explore the role of cosmic-ray pressure anisotropy, we must consider a tensor expansion of the distribution function to at least second order:
\eqb
f(\bm{x}, \bm{p}, t)=f_0(p)+\frac{\bm{p}}{p} \cdot \bm{f}_1^a(p) + \frac{5}{2}\frac{\bm{p p}}{p^2}  : {\bf S}(p)\, .
\eqe

By considering the various moments of the distribution, the physical significance of each component in the previously stated expansion is immediately apparent:

\eqb
n_{\rm cr} &=& \int d^3 p f = {4\pi}\int p^2 f_0 dp ~~~~~~~~~~~~~~~~~~~~~~~\mbox{(density) ,}\\
\bm{j}_{\rm cr} &=& e\int d^3 p  \bm{\vel} f = \frac{4\pi}{3} e\int p^2 \vel \bm{f}_1 dp~~~~~~~~~~~~~~\mbox{(current) ,}\\
{\bf P}_{\rm cr} &=& \int d^3 p \bm{\vel}\bm{p} f  = \frac{4\pi}{3}\int \vel p^3 (f_0 {\bf I} + {\bf S}) d p ~~~~~\mbox{(pressure) ,}
\eqe

where ${\bf I}$ is the unit tensor. Using the relevant orthogonality relations, it is straightforward to show that the VFP equation leads to the following system of coupled equations
\eqb
&&\label{eqn_f0}\pdiff{f_0}{t}+\bm{\nabla}\cdot(\bm{u}f_0)+\frac{{\rm v}}{3}\bm{\nabla}\cdot\bm{f_1}
=\pdiff{}{p^3}\left\lbrace p^3\left[(\bm{\nabla}\cdot\bm{u}) f_0 + {\rm S}^{ab}\pdiff{u_a}{x_b}\right]\right\rbrace\enspace,\\
&&\pdiff{{f_1^a}}{t}+u^b\pdiff{f_1^a}{x^b}
+\epsilon_{abc}\Omega^b{f_1^c}+\vel\pdiff{}{x^b}({f_0}\delta^{ab}  + {\rm S}^{ab})
 +\nu f_1^a =  \pdiff{u^b}{x^a}f_1^b
+\frac{1}{3}\pdiff{u^c}{x^c}
p^2\frac{\partial}{\partial p}\left(\frac{{f_1}^a}{p}\right) 
+\frac{1}{5}
p^2\frac{\partial}{\partial p}\left(\frac{{f_1}^b}{p}\right) \sigma^{ab} \label{eqn_f1}\enspace,\\
&&\pdiff{{\rm S}^{ab}}{t} +\pdiff{}{x^c}\left(u^c {\rm S}^{ab}\right)+\frac{\vel}{5}\Lambda^{ab} -\frac{p}{5}\pdiff{f_0}{p}\sigma^{ab}+ \left({\rm S}^{ac}\pdiff{u_b}{x_c}+ {\rm S}^{c b}\pdiff{u_a}{x_c}
-\frac{2}{3} {\rm S}^{cd}\pdiff{u_c}{x_d}\delta^{ab} \right)
-\left(\varepsilon_{acd}{\rm S}^{bc}+\varepsilon_{bcd}{\rm S}^{ac}\right)\Omega^d \nonumber \\
\label{eqn_S}
 && \enspace \enspace \enspace = \frac{5}{7}\pdiff{}{p^5}\left\lbrace p^5\left[{\rm S}^{bc}\left(\pdiff{u^a}{x^c} +\pdiff{u^c}{x^a} \right)+ 
 {\rm S}^{ac}\left(\pdiff{u^b}{x^c} +\pdiff{u^c}{x^b} \right)-\frac{2}{3}{\rm S}^{cd}\left(\pdiff{u^d}{x^c} 
 +\pdiff{u^c}{x^d} \right)\delta^{ab} +{\rm S}^{ab}\pdiff{u^c}{x^c}  \right]\right\rbrace
 - 3\nu {\rm S}^{ab}\enspace,
\eqe
where  $\bm{\Omega} = e\bm{B}/\gamma m c$ is the directional relativistic gyrofrequency, $\varepsilon_{abc}$ the Levi-Civita symbol, and summation over repeated indices is implied.
We also introduce the trace-free tensors
\eqbn
\sigma^{ab} &=& \pdiff{u^b}{x^a}+\pdiff{u^a}{x^b}-\frac{2}{3}\pdiff{u^c}{x^c}\delta^{ab} \enspace,\\
\Lambda^{ab} &=& \pdiff{f_1^a}{x^b}+\pdiff{f_1^b}{x^a}-\frac{2}{3}\pdiff{f_1^c}{x^c}\delta^{ab} \enspace,
\eqen
corresponding to the rate-of-strain tensors for the background fluid and cosmic-rays respectively. We note that these equations, using slightly different notation, have previously been derived in
\cite{Williams}, although an additional adiabatic term is included in Equation (\ref{eqn_S}) previously omitted, that ensures conservation of the trace-free nature of ${\rm S}^{ab}$.

The above equations must be solved self-consistently with the equations governing the background fluid, namely mass conservation and the magnetic induction equation (again assuming ideal MHD), together with the
cosmic-ray modified MHD momentum conservation equation
\eqb
\label{MHD_mom}
\rho\frac{d\bm{u}}{dt} &=& -\bm{\nabla} P_{\rm bg} +\frac{1}{c}\bm{j}_{\rm bg} \times \bm{B} + \eta \bm{j}_{\rm cr}  \nonumber \\
&=& -\bm{\nabla} P_{\rm bg} -\frac{c}{4\pi} \bm{B}\times(\bm{\nabla}\times\bm{B}) -\frac{1}{c}\bm{j}_{\rm cr}\times\bm{B} 
  +\eta \bm{j}_{\rm cr} \enspace,
\eqe
where we have made use of Amp\`ere's law in the last equality. The final term on the right hand side represents the collisional momentum transfer from cosmic-rays to the background, with $\eta$ an as yet to be determined
collisional transfer rate. Note that the distribution is already calculated in the local fluid frame, so it is not necessary to consider the relative drift.

Using the above definitions for cosmic-ray number density, current density and pressure, one can derive evolutionary equations for the relevant macroscopic cosmic-ray fluid quantities. 
This will ultimately require us to consider some simplifying closure relations, which we discuss in the next section.

\section{COSMIC-RAY PRESSURE AND THE FIREHOSE INSTABILITY}

From this point forward, we assume all cosmic-rays are ultra-relativistic ($p=\gamma m c$, etc.) and consist exclusively of protons. 
We note first that it is possible to modify Equation (\ref{MHD_mom}) further, by writing it in a more familiar form. Introducing the energy flux/momentum density
$$\bm{W} =\int \bm{p} f d^3p =  \frac{4\pi}{3}\int  \bm{f}_1 p^3 dp $$
and assuming our scattering rate $\nu = \Omega/h$, with $h$ (typically $\gg 1$) a momentum independent constant, it follows from Equation (\ref{eqn_f1})
\eqb
\dif{\bm{W}}{t}+(\bm{\nabla}\cdot \bm{u}) \bm{W}+(\bm{W}\cdot \bm{\nabla}) \bm{u}
=-\bm{\nabla}\,{\bf P}_{\rm cr}+\frac{1}{c}\bm{j}\times\bm{B} 
  -\eta \bm{j} \, ,
\eqe
where $\eta = |B|/ch$. Thus, restricting our attention to low frequency ($\tau \ll \Omega^{-1}, \nu^{-1}$) behaviour, we can neglect the terms on the left hand side, and one recovers the 
familiar momentum conservation equation including cosmic-ray pressure 
\eqb
\label{MHD_mom2}
\rho\frac{d\bm{u}}{dt} +\bm{\nabla} \left(P_{\rm bg}{\bf I} +{\bf P}_{\rm cr}\right) +\frac{c}{4\pi} \bm{B}\times(\bm{\nabla}\times\bm{B})=0 \, .
\eqe

We can now make use of Equations (\ref{eqn_f0}) and (\ref{eqn_S}) to determine the form of the anisotropic pressure tensor. 
Defining the rank 3 tensor
\eqbn
Q^{abc} &=& \frac{1}{m}\int p^a p^b p^c f \frac{d^3 p}{p^0} =  \left\lbrace
\begin{array}{l c} 
\frac{4\pi}{3m}\int dp~ p^4  ~ \left(f_0\delta^{bc} + S^{bc}\right) &   a=0, ~~b,c > 0 \\
\frac{4\pi}{15m} \int  \beta p^4 \left[f_1^a\delta^{bc}+f_1^b\delta^{ca}+f_1^c\delta^{ab}\right]dp &  a,b,c > 0
\end{array}\right. \, ,
\eqen
it follows that in the ultra-relativistic limit
\eqb
\frac{d Q^{0ab}}{dt} + c\pdiff{Q^{abc}}{x^c} + Q^{0ab}\pdiff{u^c}{x^c}+ Q^{0cb}\pdiff{u^a}{x^c}+ Q^{0ac}\pdiff{u^b}{x^c}
-\frac{eB^d}{mc}\left(\epsilon_{acd}P^{bc}+\epsilon_{bcd}P^{ac}\right)=-3\bar{\nu}\Pi^{ab}\, ,
\eqe
where $\Pi^{ab}$ is the trace free part of the cosmic-ray pressure tensor, and we have taken advantage of the fact that $\bar{\nu}=eB/hmc$.
We note that, for a $f\propto p^{-4}$ spectrum with range $p_1<p<p_2$ it follows that $Q^{0ab}/P^{ab} \sim (p_2/mc) \log(p_2/p_1)$, while $Q^{abc}/P^{ab}$ is typically smaller by a fraction
$u_{\rm sh}/c$\footnote{
Formally speaking the ratio of the second to third term is, in the diffusion approximation $\sim (u_{\rm sh}/c)^2 (\nu/k \delta u)$. Assuming $\delta u \sim v_{\rm A}$, it follows that the heat flow is negligible
on scales $k^{-1} \ll (c/u_{\rm sh}) M_{\rm A} \langle \lambda_{\rm mfp}\rangle$, ie. the shock crossing time is less than the time taken for an Alfv\'en wave to transmit information across the precursor scaleheight.
}
. Hence, in the same limit as before ($\tau \ll \Omega^{-1}, \nu^{-1}$ at the upper energy range), the leading terms in this equation are
$$ \left(\epsilon_{acd}P^{bc}+\epsilon_{bcd}P^{ac}\right)B^d = 0\enspace .$$
This equation is satisfied by any tensor of the form
$${\bf P}_0= p_\|{\bf bb} + p_\bot(\bf{I}-{\bf bb}) \, ,$$
where ${\bf b}=\bm{B}/B$ is the unit vector along the field, and $p_\|, ~p_\bot$ as yet undetermined constants.
In order to determine $p_\|, ~p_\bot$, we make the following approximations. We neglect collisions and the heat flux term (these can be checked \emph{a posteriori}), and define a slowly varying weighted Lorentz factor
\eqb
\langle \gamma \rangle = \frac{\int  \gamma \vel^a p^b f d^3p}{\int \vel^a p^b f d^3p} =  \frac{Q^{0ab}}{P^{ab}}\, ,
\eqe
allowing us to write, to next leading order (see also \cite{Kulsrud})
\eqb
\frac{d P_0^{ab}}{dt} + P_0^{ab}\pdiff{u^c}{x^c}+ P_0^{cb}\pdiff{u^a}{x^c}+ P_0^{ac}\pdiff{u^b}{x^c}
=\left(\epsilon_{acd}P_1^{bc}+\epsilon_{bcd}P_1^{ac}\right)\tilde{\Omega}^d~,
\eqe
where $\tilde{\bm{\Omega}} = e\bm{B}/\langle \gamma \rangle m c$. 
Clearly, the right hand side of this equation is trace free, and similarly it vanishes by contraction on ${\rm b}_a {\rm b}_b$. 

Using $$\bm{\nabla}\cdot\bm{u} = -\frac{1}{\rho}\frac{d\rho}{dt}
\mbox{~~~and~~~} \frac{dB}{dt} = {\bf b}\cdot \frac{d\bm{B}}{dt} =  {\bf b}\cdot\left[(\bm{B}\cdot\bm{\nabla})\bm{u}-\bm{B}(\bm{\nabla}\cdot\bm{u})\right]~,$$
the standard double adiabatic equations follow:
\eqb
\frac{d}{dt}\left(\frac{p_\|B^2}{\rho^3}\right) = \frac{d}{dt}\left(\frac{p_\bot}{\rho B}\right) = 0 \, .
\eqe

This closes our system of equations, which are now in exactly the form that reproduces the well-known result for parallel modes (see for example \cite{kralltrivelpiece}):
\eqb
\omega^2 = \frac{k^2}{\rho}\left[\frac{B^2}{4\pi}+p_\bot-p_\|\right] \, ,
\eqe
which is purely growing in the limit 
\eqb
\label{FH_condition}
p_\|-p_\bot > \frac{B^2}{4\pi} \, .
\eqe

\section{FIREHOSE IN SNR PRECURSORS}

While we have identified the necessary conditions for the onset of firehose instability, the discussion up to this point has not made any connection to actual supernovae,
and specifically what physical values $p_\|$ and $p_\bot$ might take. For simplicity we consider a planar steady shock with velocity $u_{\rm sh}$, with magnetic field and shock normal along the $x$-axis.
Again, using Equations (\ref{eqn_f0})-(\ref{eqn_S}), the steady state solution in the upstream plasma has 
\eqb
f_1^1 = 3 \frac{u_{\rm sh}}{c} f_0\enspace,  \enspace\enspace\enspace S^{11} = -2 S^{22} = -2S^{33} = \frac{4}{5}\left(\frac{u_{\rm sh}}{c}\right)^2 f_0 \, .
\eqe

Using these numbers, the firehose condition, Equation (\ref{FH_condition}) can be expressed as
\eqb
\frac{6}{5}\left(\frac{u_{\rm sh}}{c}\right)^2 M^2_{\rm A} \left( \frac{P^0_{\rm cr}}{\rho u_{\rm sh}^2} \right) > 1 \, ,
\eqe
where $P_{\rm cr}^0$ is the isotropic cosmic-ray scalar pressure and $M_{\rm A}$ the Alfv\'en Mach number of the shock. We also recall, that in the previous section, we neglected the 
role of collisions and heat flux in the pressure tensor equation. The first of these approximations is clearly justified, provided $\nu \ll \Omega$ ($h >> 1$), i.e. scattering is far from the Bohm limit. 
The latter approximation, of negligible heat flux is more controversial \cite{BlandfordFunk}, although it generally valid provided the shock velocity remains non-relativistic.

Assuming the above conditions are satisfied, the corresponding growth rate is 
\eqb
\label{FH}
\Gamma_{\rm FH} \sim ku_{\rm sh}\left[  \left(\frac{u_{\rm sh}}{c}\right)^2 \left( \frac{P_{\rm cr}}{\rho u_{\rm sh}^2}\right)\right]^{1/2} \, .
\eqe

It is interesting to note that the growth rate is very similar to that of the \emph{strongly modified} resonant ion-cyclotron instability, \cite{Bell04}
\eqb
\label{IC}
\Gamma_{\rm  IC} \approx ku_{\rm sh}\left[ \frac{1}{\ln(p_{\rm max}/p_{\rm min})}\left(\frac{u_{\rm sh}}{c}\right) \left( \frac{P_{\rm cr}}{\rho u_{\rm sh}^2}\right)\right]^{1/2}\, .
\eqe
We note that the condition for strong modification is $M^2_{\rm A} ({u_{\rm sh}}/{c}) ( {P^0_{\rm cr}}/{\rho u_{\rm sh}^2} ) / \ln(p_{\rm max}/p_{\rm min}) > 1$, which is 
also not dissimilar to the firehose condition.
However, while the growth rates are comparable, the firehose instability is purely growing, the real and imaginary parts of the frequency ion-cyclotron are comparable in the strongly modified case, 
and is dominated by the real part in the unmodified case. Additionally, the ion-cyclotron instability is only expected to grow for waves with polarisation in the same sense as the cosmic-rays' 
zeroth order helical motion in the mean field, 
while the firehose, being non-resonant, does not depend on the sense of rotation,  and is thus unstable to both polarisations, or indeed in the presence of coherent curved magnetic field structure. 

\begin{figure} 
  \centerline{\includegraphics[width=150pt, angle=-90]{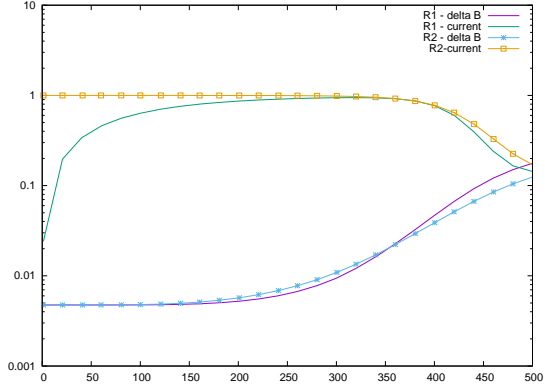}}
 
  \caption{Growth of magnetic field fluctuations and driving current as a function of time from 3D simulations. Magnetic field is normalised to the initial mean field, 
  while the cosmic ray current is normalised to its value in the diffusive approximation. Time is in units of $\Omega_g^{-1}$.}
  
\end{figure}

\section{HYBRID MHD-VFP SIMULATIONS}

We present here some preliminary simulations to explore the non-linear behaviour of cosmic-rays interacting with an MHD plasma on large scales. 
The simulations were performed in 3D, using Equations (\ref{eqn_f0})- (\ref{MHD_mom}) together with the equations for 
mass and energy conservation, and the magnetic induction equation. We consider a periodic domain, with magnetic field along the $x$-axis, and background fluid 
initially at rest. To minimise numerical memory requirements, we employ the same technique 
used in \cite{RevilleBell}, and solve the VFP equations for a single particle momentum $p_0$, replacing all momentum derivatives assuming a $p^{-4}$ power-law. This technique 
is valid provided the energy changes are small, but is essential to capture the important $E \times B$ drifts with respect to the background fluid.  

Figure 1 shows the evolution of the cosmic-ray current and magnetic field fluctuations as a function of time for two different simulations, R1 \& R2,
both of which satisfy the firehose condition. Simulations R1 had a cosmic-ray pressure $P_{\rm cr}/\rho u_{\rm sh}^2 = 0.02$ while this number is $0.01$ for R2. The shock velocity and 
collision frequency in both simulations were $0.1 c$ and $\Omega_{\rm g}/100$ respectively. A grid resolution of $\Delta x = r_{\rm g,0}/5$ was used, where $r_{\rm g,0}$ is the gyroradius in the mean field.
Both simulations included a uniform external driving term (see \cite{RevilleBell}), which proves to be essential, 
since any initial anisotropy would be damped on a timescale $\nu^{-1}$. To explore different possibilities, R1 starts from rest ($f(t=0)$ is isotropic), while R2 is initialised with the diffusive solution given above. 

Since both the firehose and ion-cyclotron instability have a growth rate proportional to $k$, the fastest growing mode appears to be occurring close to the grid scale. 
The growth rate of the field is consistent with either Equation (\ref{FH}) and (\ref{IC}) with wavelength $\lambda \sim 2 r_g$. The polarisation appears to vary with position of the line-out in the $y-z$ plane 
making it ambiguous as to which instability is dominating.  
However, the modes appear to be purely growing which is more characteristic of the firehose instability. It has been previously suggested by \cite{RevilleBell} that sub-Larmor scattering 
of the cosmic-rays allows them to decouple from long wavelength fluctuations, and allow purely growing modes. However, this is unlikely to be the case here as there is insufficient structure 
below the Larmor scale. 

Finally we note, in both cases, the cosmic-ray current is rapidly damped when $\delta B/B_0$ exceeds a few percent level. Given that $\nu = \Omega_{\rm g}/100$ it is expected at about this level that gyration in
the non-uniform fields dominates over the imposed small angle scattering. Future simulations can alter the driving to maintain a steady current.

\section{CONCLUSIONS}

We have identified the minimal conditions for the cosmic-ray driven firehose to occur in SNR precursors, in particular with regards the necessary approximations. As has been previously pointed out 
by \cite{BlandfordFunk}, the biggest limitation concerning these approximations may well be the neglect of the heat flux, which for fast shocks ($>0.1 c$) can be comparable to other first order terms. 
We note that we have only considered the case of cosmic-ray current and pressure 
anisotropy driven by a large scale gradient, using the so-called diffusive approximation. If scattering is weak, and particles can escape the accelerator more freely, it is conceivable that the pressure anisotropy
takes a different form. However, how to implement such an effect without performing full shock simulations is not obvious.   

Preliminary simulations indicate that , irrespective of the instability operating, a significant reduction in the mean free path, or equivalently, an enhancement of the CR confinement, can be achieved in a relatively modest number of Larmor periods ($\sim 10-50$ for our chosen simulation parameters). If for example, one considers a very young SNR shock in a highly magnetised but dense plasma (such that the Alfv\'en Mach number is still large), there is a clear prospect for self confinement on timescales that are consistent with the expansion time of the remnant, even at PeV energies.



\section{ACKNOWLEDGMENTS}
BR gratefully acknowledges valuable discussions with Tony Bell and John Kirk.

\nocite{*}
\bibliographystyle{aipnum-cp}%
\bibliography{Gamma_Proceedings}%

\end{document}